\documentclass[a4paper,11pt]{article}
\usepackage{jcappub} 
\usepackage{lineno}

\usepackage{graphicx}
\usepackage{dcolumn}
\usepackage{bm}
\usepackage{amssymb}
\usepackage{latexsym}
\usepackage{booktabs}
\usepackage{amsmath}
\usepackage{multirow}
\usepackage{url}
\usepackage{hyperref}
\usepackage{footnote}
\usepackage{float}
\usepackage{threeparttable}
\usepackage{color}
\usepackage{array}
\usepackage{enumerate}
\usepackage{adjustbox}
\usepackage{BOONDOX-cal}

\usepackage{subcaption}
\usepackage{makecell}
\usepackage{diagbox}
\usepackage{epstopdf}
\usepackage{epsfig}
\usepackage{longtable}
\usepackage{supertabular}
\usepackage{algorithm}
\usepackage{pifont}
\usepackage{algorithmic}
\usepackage{changepage}
\usepackage{setspace}

\newcommand{\blue}[1]{\textcolor{blue}{#1}}

\begin{document}	

	\title{Prospects for cosmological research with the FAST array: 21-cm intensity mapping survey observation strategies}

	\author[a]{Jun-Da Pan,}
        \author[a]{Peng-Ju Wu,}
	\author[a]{Guo-Hong Du,}
        \author[a]{Yichao Li}
	\author[a,b,c,\href{note}{\ast}]{and Xin Zhang\note[$\blue \ast$]{Corresponding author.}}

	\affiliation[a]{Liaoning Key Laboratory of Cosmology and Astrophysics, College of Sciences, Northeastern University, Shenyang 110819, China}
	\affiliation[b]{MOE Key Laboratory of Data Analytics and Optimization for Smart Industry, Northeastern University, Shenyang 110819, China}
	\affiliation[c]{National Frontiers Science Center for Industrial Intelligence and Systems Optimization, Northeastern University, Shenyang 110819, China}

	\emailAdd{panjunda@stumail.neu.edu.cn}
	\emailAdd{wupengju@stumail.neu.edu.cn}
        \emailAdd{duguohong@stumail.neu.edu.cn}
	\emailAdd{liyichao@mail.neu.edu.cn}
	\emailAdd{zhangxin@mail.neu.edu.cn}

	\abstract{Precise cosmological measurements are essential for understanding the evolution of the universe and the nature of dark energy. The Five-hundred-meter Aperture Spherical Telescope (FAST), the most sensitive single-dish radio telescope, has the potential to provide the precise cosmological measurements through neutral hydrogen 21 cm intensity mapping sky survey. This paper primarily explores the potential of technological upgrades for FAST in cosmology. The most crucial upgrade begins with equipping FAST with a wide-band receiver ($0 < z < 2.5$). This upgrade can enable FAST to achieve higher precision in cosmological parameter estimation than the Square Kilometre Array Phase 1 Mid-Frequency Array. On this basis, expanding to a FAST array (FASTA) consisting of six identical FASTs would offer significant improvements in precision compared to FAST. Additionally, compared with the current results from the data combination of cosmic microwave background, baryon acoustic oscillations (optical galaxy surveys), and type Ia supernovae, FASTA can provide comparable constraints. Specifically, for the dark-energy equation-of-state parameters, FASTA can achieve $\sigma(w_0) = 0.09$ and $\sigma(w_a) = 0.33$.}
	\maketitle
	\section{Introduction}

    Dark energy is a component of the universe that causes the late-time cosmic acceleration. Although it is known to comprise 68\% of the total energy density of the universe, the nature of dark energy remains unknown. The $\Lambda$ cold dark matter ($\Lambda$CDM) model, the standard model of cosmology, treats dark energy as a constant energy density filling space homogeneously. {However, recent observations of baryon acoustic oscillations (BAO) from the Dark Energy Spectroscopic Instrument (DESI) have shown deviations from the $\Lambda$CDM model \cite{aghamousa2016desi, adame2024desi, adame2024desi2, Craig:2024tky, Green:2024xbb, Wu:2024faw, Du:2024pai, Li:2024qso}. The result suggests that dark energy may be a dynamical energy density evolving over time or indicate the presence of other new physics. Nonetheless, further observations and advanced instruments are needed to confirm these possibilities and to better understand the evolutionary dynamics of dark energy.}

BAO from the large-scale structure (LSS) is a late-time cosmological probe that can be used as a standard ruler to study the expansion history of the universe \cite{blake2003, Cole2005, Percival2010, Beutler2011, Blake2011, Ross2015, Alam2017, Ata2018}. Currently, BAO studies primarily rely on optical galaxy surveys \cite{york2000, percival2001, kaiser2002pan, dark2016dark, aghamousa2016desi}. While effective, these surveys are limited in terms of efficiency and coverage.  To overcome these limitations and achieve more precise cosmological measurements, advancements in radio astronomy have developed a promising method known as 21 cm intensity mapping (21 cm IM). After reionization, neutral hydrogen ($\mathrm{H\ {\scriptstyle I}}$) mainly exists within galaxies. This technique measures the intensity of the characteristic 21 cm emission line from $\mathrm{H\ {\scriptstyle I}}$ to study the LSS \cite{furlanetto2006, chang2010, switzer2013determination, masui2013measurement, pritchard201221}. This method does not require resolving individual galaxies, thus it has high survey efficiency. The characteristics of the 21 cm signal allow us to probe deeper into the universe and directly obtain redshift information. Recent forecasts and observational studies have confirmed the potential of 21 cm IM in cosmology \cite{bull2015late, Zhang:2019dyq, Zhang:2019ipd, Jin:2020hmc,hu2020forecast, Zhang:2021yof, Jin:2021pcv, wu2022prospects, Wu:2022dgy, wu2023prospects, chen2023joint, Zhang:2023gaz, Li:2023zer, cunnington2023h}. {In addition to the 21 cm line, intensity mapping of other spectral lines, such as CO and CII, has also attracted attention. These lines can provide information about the LSS and the distribution of molecular and ionized gas in galaxies \cite{2019PhRvL.123y1301B, 2019PhRvD.100l3522B, 2024PhRvL.132x1001F}.}\par

	To achieve more precise cosmological measurements, advanced radio telescopes are utilized in 21 cm IM. These telescopes include those currently in use, such as the Five-hundred-meter Aperture Spherical Telescope (FAST) \cite{nan2011fast}, the MeerKAT \cite{2017arXiv170906099S,2021MNRAS.505.3698W,2020MNRAS.tmp.3636L}, the Tianlai \cite{chen2011radio,chen2012tianlai,2020SCPMA..6329862L,2021MNRAS.506.3455W,2022MNRAS.517.4637P,2022RAA....22f5020S}, the Canadian Hydrogen Intensity Mapping Experiment (CHIME) \cite{Newburgh:2014toa}. {Currently, 21 cm signals have been observed through cross-correlation with galaxy surveys \cite{chang2010, masui2013measurement, Anderson:2017ert, 2020MNRAS.tmp.3636L, Tramonte_2020, Wolz_2021, 2023HII}, and the MeerKAT telescope has also detected the autocorrelation spectrum on Mpc scales \cite{Paul:2023yrr}. However, the large scale autocorrelation power spectrum still has not been observed \cite{switzer2013determination}.
}Future instruments planned for use include the Square Kilometre Array (SKA) \cite{dewdney2009ska, santos2015cosmology, bourke2015advancing, 2020, an2022status}, the Baryon Acoustic Oscillations from Integrated Neutral Gas Observations (BINGO) \cite{battye2012bingo, wuensche2019bingo}, the Hydrogen Intensity and Real-time Analysis eXperiment (HIRAX) \cite{Newburgh:2016mwi}, and {the Canadian Hydrogen Observatory and Radio-transient Detector (CHORD)} \cite{2019clrp.2020...28V}. These advanced facilities are expected to further enhance the capabilities of 21 cm IM, enabling more detailed and accurate cosmological studies.

    Among these, FAST is particularly significant due to its large aperture and sensitivity. The current FAST has achieved substantial success in many areas, particularly in fast radio burst and gravitational wave researches \cite{Li_2021, Niu_2022, Xu_2023}. However, to harness its potential for 21 cm IM and to extend its capability for cosmological studies, several upgrades are planned for FAST  \cite{zhang2023performance, xue2023pulsar, yin2023pose,jiangpeng}. 
    
    This paper aims to explore how potential future upgrades to FAST could promote cosmological research by using 21 cm IM. For the expansion of FAST itself, equipping it with a wide-band receiver aims to extend its 21 cm signal detection capability, enabling broader redshift coverage \cite{zhang2023performance}. This will allow us to observe a larger volume of the universe, aiding in constructing a more precise three-dimensional map of the LSS. Moreover, a phased array feed (PAF) is planned to enhance its survey efficiency \cite{landon2010phased, hotan2021australian}.
    Additionally, the expansion of the construction around FAST is also considered. The plan involves building 40-meter aperture radio telescopes (40-m antennas) in two phases around FAST: the first phase involves constructing six 40-m antennas, and the second phase will increase this number to twenty-four \cite{jiangpeng}. 
	The most significant upgrade is the FAST array (FASTA), which plans to add five more identical 500-meter diameter telescopes to join the current FAST \cite{xue2023pulsar}. This network will significantly improve sensitivity and survey efficiency, enhancing the ability to explore LSS and providing higher precision in cosmological parameter estimation.
	
    In this paper, we present a forecast for estimating cosmological parameters using simulated 21 cm IM data and discuss how the various extensions of FAST will improve the precision of these constraints.
    
    In Section~\ref{sec2}, we introduce the 21 cm IM simulation and the methods used to constrain cosmological parameters. In Section~\ref{sec3}, we discuss the performance of various configurations for 21 cm IM and the resulting cosmological constraints. In Section~\ref{sec4}, we conclude with perspectives for future research.
	
	In our analysis, we adopt the \textit{Planck} best-fit \(\Lambda\)CDM model as fiducial cosmology  \cite{planck2018}. The parameters are \(H_0 = 67.3\) km s\(^{-1}\) Mpc\(^{-1}\), \(\Omega_\mathrm{m} = 0.317\), \(\Omega_\mathrm{b} = 0.0495\), \(\Omega_K= 0\), \(\sigma_8 = 0.812\), and \(n_\mathrm{s} = 0.965\).

\section{21 cm intensity mapping}\label{sec2}
   \subsection{Signal}\label{sec2.1}
   21 cm IM measures the intensity of 21 cm signals and converts it into brightness temperature.  The mean temperature of $\mathrm{H\ {\scriptstyle I}}$ can be expressed as \cite{bull2015late}
   \begin{equation}
   	\overline{T}_\mathrm{b} = \frac{3}{32 \pi} \frac{\mathcal{h} c^3 A_{10}}{k_\mathrm{B} m_\mathrm{p} \nu_{21}^2} \frac{(1+z)^2}{H(z)} \Omega_{\mathrm{H \,{\scriptscriptstyle I}}}(z) \rho_{\mathrm{c}, 0} \ ,
   \end{equation}
   where $\mathcal{h}$ is the Planck constant, \( c \) is the speed of light, \( A_{10} \) is the Einstein coefficient for spontaneous emission \cite{furlanetto2006}, \( k_\mathrm{B} \) is the Boltzmann constant, \( m_\mathrm{p} \) is the proton mass, \( \nu_{21} \) is the $\mathrm{H\ {\scriptstyle I}}$ line frequency, \( H(z) \) is the Hubble parameter, \( \Omega_{\mathrm{H\,{\scriptscriptstyle I}}}(z) \) is the fractional density of $\mathrm{H\ {\scriptstyle I}}$, and \( \rho_{\mathrm{c}, 0} \) is the current critical density of the universe.
   
    The signal covariance is expressed in terms of the power spectrum
    \cite{blake2003,bull2015late}
		\begin{equation}
		C^\mathrm{S}(\boldsymbol{q}, y) = \frac{T_b^2(z_i)}{r^2 r_\nu} (b_{\mathrm{H \,{\scriptscriptstyle I}}} + f\mu^2)^2 \exp(-k^2\mu^2\sigma_{\mathrm{NL}}^2)  P(k, z),
	\end{equation}
	where \( \boldsymbol{q} = \boldsymbol{k}_{\perp} r \) and \( y = k_{\|} r_\nu \). \( r \) denotes the comoving distance to \( z \), and \( r_\nu \) corresponds to the comoving distance for the frequency interval. \( b_\mathrm{H \,{\scriptscriptstyle I}} \) represents the bias factor of $\mathrm{H\ {\scriptstyle I}}$ \cite{Xu_2014}. \( f \) is the linear growth rate of the structure. \( \mu = k_{\|} / k \) is the cosine of the angle between the wavevector \( \boldsymbol{k} \) and the line of sight. \( \sigma_{\mathrm{NL}}=7\ \mathrm{Mpc} \) represents the non-linear dispersion scale \cite{Li_2007}. \( P(k, z) \) is the matter power spectrum at wavenumber \( k \) and redshift \( z \), which can be calculated by {\tt CAMB} \cite{Lewis_2000}.

	\subsection{Noise}\label{sec2.2}
	The noise covariance can be written as \cite{bull2015late}
	\begin{equation}
		C^\mathrm{N}(\boldsymbol{q}, y)=\frac{T_{\text {sys }}^2}{t_{\text {tot }} \Delta \nu \ N_\mathrm{b} N_\mathrm{d} } U_{\text {bin }}B_{\perp}^{-2} B_{\|}^{-1},
	\end{equation}
	where \( T_{\mathrm{sys}} \) includes the instrumental temperature \( T_{\mathrm{inst}} \) and the sky background radiation \( T_{\text{sky}} \approx 60 \mathrm{~K} \times (\nu / 300\  \mathrm{MHz})^{-2.5} \).  \( t_{\mathrm{tot}} \) represents the total integration time, \( N_\mathrm{b} \) is the number of beams, \( N_\mathrm{d} \) is the number of dishes in the array, and \( \Delta\nu \) is the total bandwidth. \( U_{\text{bin}} = S_{\text{area}} \Delta \tilde{\nu} \) represents the survey volume at a given redshift, where \( S_{\text{area}} \) is the survey area and \(\Delta \tilde{\nu} = \Delta{\nu}/\nu_{21}\) is the dimensionless bandwidth for a given redshift bin. \( B_{\|} \) is the radial beam response
	\begin{equation}
		B_{\|}(y)=\exp \left(-\frac{\left(y \delta \nu / \nu_{21}\right)^2}{16 \ln 2}\right),
	\end{equation}
	where \( \delta\nu \) is the bandwidth of an individual frequency channel. \( B_{\perp} \) is the transverse beam response in single dish mode
	\begin{equation}
		B_{\perp}(\boldsymbol{q})=\exp \left(-\frac{\left(q \frac{\lambda}{D_\mathrm{dish}}\right)^2}{16 \ln 2}\right),
	\end{equation}
	where \( D_\mathrm{dish} \) is the diameter of the dish.

	\subsection{Foreground}\label{sec2.3}
    The foreground is an important component of the signal received by the telescope. It primarily originates from our own galaxy and extragalactic point sources. At this stage, the effectiveness of methods for subtracting foregrounds from observational data has been very limited. {The methods include Polynomial Fitting, Principal Component Analysis, Independent Component Analysis, and others \cite{Wang_2006, Morales_2012, Parsons_2012, Liu_2014, Shaw_2015, Zhu_2018, zuo201821, Carucci_2020, Cunnington_2021, ni2022eliminating, Gao:2022xdb, foreground}.} We assume a method that can effectively subtract the foreground, and the residual foreground model can be written as \cite{bull2015late, Santos_2005}

	\begin{equation}
		C^{\mathrm{F}}(\boldsymbol{q}, y)=\varepsilon_{\mathrm{FG}}^2 \sum_X A_X\left(\frac{l_p}{2 \pi q}\right)^{n_X}\left(\frac{\nu_p}{\nu_i}\right)^{m_X},
	\end{equation}
	where \(\varepsilon_{\mathrm{FG}}\) is a factor that describes the efficiency of foreground subtraction. \(\varepsilon_{\mathrm{FG}} = 1\) represents no foreground subtraction, while \(\varepsilon_{\mathrm{FG}} = 0\) represents perfect foreground subtraction. We assume \(\varepsilon_{\mathrm{FG}}=10^{-6}\), as this level is effective for extracting cosmological signals. $X$ is the varies foreground component shown in Table \ref{table:foreground_parameters}. 

\begin{table}[H]
    \small
    \captionsetup{justification=raggedright, singlelinecheck=false}
	\caption{Foreground model parameters at \( l_p = 1000 \) and \( \nu_p = 130 \) MHz \cite{Santos_2005}.}
	\centering
	\begin{tabular}{lccc}
		\hline\hline
		Foreground & \( A_X \) [mK\(^2\)] & \( n_X \) & \( m_X \) \\
		\hline
		Extragalactic point sources & 57.0 & 1.1 & 2.07 \\
		Extragalactic free–free & 0.014 & 1.0 & 2.10 \\
		Galactic synchrotron & 700 & 2.4 & 2.80 \\
		Galactic free–free & 0.088 & 3.0 & 2.15 \\
		\hline
	\end{tabular}
	\label{table:foreground_parameters}
\end{table}

\subsection{Instrument parameters}\label{sec2.4}
In this paper, we consider several different configurations of FAST and the Square Kilometre Array Phase 1 Mid-Frequency Array (SKA1-Mid) for simulating the 21 cm IM. The specific instrument parameters are listed in Table \ref{table:instrument}.

The current capability of FAST to detect 21 cm signals covers the redshift range \( 0 < z < 0.35 \). A new wide-band receiver under test extends this range to \( 0 < z < 1.84 \) \cite{zhang2023performance}. 
Currently, DESI measurements have mapped the cosmological distance-redshift relation with an accuracy of 1\%–3\% across redshifts below $z<2.5$ \cite{adame2024desi}. This paper assumes that the upgraded new wide-band receiver on FAST can detect 21 cm signals up to \(0 < z < 2.5 \). 

Additionally, constructing 6$\times$40-m antennas or 24$\times$40-m antennas around FAST is considered. These dishes will use the same receiver and feed system as FAST, so their frequency bands and $N_\mathrm{b}$ will be identical. Given the outstanding performance of FAST, we also consider constructing five additional identical FAST telescopes to form FASTA, significantly improving survey efficiency. Furthermore, we consider equipping FAST with a PAF to increase the value of $N_\mathrm{b}$ to 100.

Regarding instrument temperature, in Ref.~\cite{Jiang_2020} the authors measured the temperature of FAST near boresight to be 20 K in the L-band. We assume that the frequency band can be broadened while maintaining the instrument temperature at 20 K. For the PAF, we assume the temperature to be approximately 30 K due to the increase in the number of beams. The total observation time is assumed to be 10000 hours. The calculation for the survey area is as follows

	\begin{equation}
		S_{\text{area}} = \eta  \int_{\phi} \int_{\theta} \sin \theta \mathrm{d} \theta \mathrm{d} \phi \ ,
	\end{equation}
	where \(\eta\) represents the fraction of the initial survey area that remains usable for scientific observation after excluding the Milky Way. For FAST/FASTA, we consider \(\eta = 0.7\) and the range of declination from $-14^\circ$ to $+66^\circ$, yielding an area of approximately 8100 $\text{deg}^2$.

For SKA1-Mid, we consider Band 1 ($0.35<z<3$) planned for 21 cm IM, with the survey area of 20000 deg\(^2\) and the total observation time of 10000 hours. {The instrument temperature is modeled based on actual measurements and simulation data from the SKA Observatory, and the form is as follows \cite{2020, braun2017anticipated}}
\begin{equation}
T_{\mathrm{inst}}=15 \mathrm{~K}+30 \mathrm{~K}\left(\frac{f}{\mathrm{GHz}}-0.75\right)^2 .
\end{equation}

\begin{table*}[htbp]
 \captionsetup{justification=raggedright, singlelinecheck=false}
 \caption{Instrument parameters for the 21 cm IM forecast.}
 \centering

 \setlength{\tabcolsep}{2mm}
 \small
 \renewcommand{\arraystretch}{1.3}
\fontsize{11pt}{12pt}\selectfont
  \begin{tabular}{lcccccccc}
   \hline\hline
   Instrument & $z_{\text{min}}$ & $z_{\text{max}}$ & $N_\mathrm{d}$ & $N_\mathrm{b}$ & $D_\mathrm{dish}$ [m] & $S_{\text{area}}$ [deg$^2$] & $t_{\text{tot}}$ [h] & $T_{\text{inst}}$ [K] \\
   \hline
   FAST & 0 & 2.50 & 1 & 19 & 300 & 8100 & 10000 & 20 \\
   SKA1-Mid & 0.35 & 3.00 & 197 & 1 & 15 & 20000 & 10000 & 18-28 \\
   40-m antennas & 0 & 2.50 & 6 or 24 & 19 & 40 & 8100 & 10000 & 20 \\
   FASTA & 0 & 2.50 & 6 & 19 & 300 & 8100 & 10000 & 20 \\
   FAST(PAF) & 0 & 2.50 & 1 & 100 & 300 & 8100 & 10000 & 30 \\
   \hline
  \end{tabular}
 \label{table:instrument}
\end{table*}
	\subsection{The Fisher matrix}\label{sec2.5}
	To achieve optimal observational results, different telescopes require different survey strategies, so the methods for calculating \(C^{\mathrm{tot}}\) also vary. For a single dish or identical dishes observing the same area of the sky (such as FAST or FASTA), the total covariance matrix is expressed as
\begin{equation}
C^{\mathrm{tot}} = C^{\mathrm{S}} + C^{\mathrm{N}} + C^{\mathrm{F}}.
\end{equation}

For the combination of different telescopes, we consider two scenarios. The first scenario involves two telescopes in close geographic proximity observing the same area of the sky, such as combining 40-m antennas and FAST. This approach maximizes the reduction of thermal noise impact on observations. Since \(C^{\mathrm{S}}\) and \(C^{\mathrm{F}}\) are only related to the observed sky area, they remain fixed. Different telescopes have varying \(C^{\mathrm{N}}\) due to their differing thermal noise integration effects. Therefore, the total covariance matrix can be expressed as
\begin{equation}
C^{\mathrm{tot}} = C^{\mathrm{S}} + \left((C_{1}^{\mathrm{N}})^{-1} + (C_{2}^{\mathrm{N}})^{-1}\right)^{-1} + C^{\mathrm{F}},
\end{equation}
where \(C_{1}^{\mathrm{N}}\) and \(C_{2}^{\mathrm{N}}\) are the noise covariance matrices of the different dishes, with \((C^{\mathrm{N}})^{-1}\) denoting their inverse matrices. The second scenario involves two telescopes with significant differences in latitude, resulting in a very small common observable sky area. Therefore, they must observe different areas of the sky, which also maximizes the survey volume. Since both the observed sky areas and the telescopes are different, the total covariance matrix is given by
\begin{equation}
C^{\mathrm{tot}} = \left((C_{1}^{\mathrm{tot}})^{-1} + (C_{2}^{\mathrm{tot}})^{-1}\right)^{-1},
\end{equation}
where \(C_{1}^{\mathrm{tot}}\) and \(C_{2}^{\mathrm{tot}}\) are the total covariance matrices of the different telescopes.

	After obtaining the total {covariance} matrix \(C^{\mathrm{tot}}\), we can transform it into the Fisher matrix of observables \(\{p_{i}\}\) \cite{bull2015late}
	\begin{equation}
		F_{i j} = \frac{1}{2} U_{\text{bin}} \int \frac{\mathrm{d}^2 q \, \mathrm{d} y}{(2 \pi)^3} \left[\frac{\partial \ln C^\mathrm{tot}(\boldsymbol{q}, y)}{\partial p_{i}} \frac{\partial \ln C^\mathrm{tot}(\boldsymbol{q}, y)}{\partial p_{j}}\right].
	\end{equation}
	In this paper, we mainly calculate the Fisher matrix for several key observables to constain the evolution of LSS, as well as the growth and geometry of the universe. These include the transverse distance indicator $D_\mathrm{A}(z)$, radial distance indicator $H(z)$ and the parameter combination $f\sigma_{8}(z)$ from redshift space distortion (RSD), where \( f(z) \) is growth rate and \(\sigma_8\) is the amplitude of matter density fluctuations on a scale of \(8 h^{-1}\) Mpc (with $h$ the dimensionless Hubble constant).

	\subsection{Cosmological parameter constraints}\label{sec2.6}
    We use the Fisher matrix to propagate errors from observational errors to the BAO parameters $D_\mathrm{A}(z)$, $H(z)$, and $f\sigma_8(z)$. Then, we use the errors of BAO parameters to form the likelihood function for cosmological models. Finally, we employ Markov Chain Monte Carlo (MCMC) analysis to constrain the cosmological parameters \cite{Witzemann_2018}. {Our considered parameter set is $\{ H_0, \Omega_\mathrm{m}, \Omega_\mathrm{b}, \sigma_8, n_\mathrm{s}, w_{0}, w_{a} \}$. We assume a flat universe with $\Omega_K=0$  and fix the total neutrino mass $\sum m_\nu=\mathrm{0.06\ eV}$. For the free parameters, we assume the priors within the following ranges: $H_0 \in [50, 80]$ km s$^{-1}$ Mpc$^{-1}$, $\Omega_\mathrm{m} \in [0, 0.7]$, $\Omega_\mathrm{b}  \in [0.03, 0.07]$, $\sigma_8 \in [0, 2]$, $n_\mathrm{s} \in [0.8, 1.2]$, $w_{0} \in [-3, 5]$, and $w_{a} \in [-5, 5]$. Although BAO alone cannot constrain \( n_\mathrm{s} \) and \( \Omega_\mathrm{b} \), we consider them as free parameters to maintain consistency of the parameter space with the cases when considering CMB and supernova (SN) data.}

 To compare with the mainstream observations, we calculated the constraints from the latest observations. The data used include: CMB -- \textit{Planck} 2018 high-\(\ell\) \textit{TT}, \textit{TE}, \textit{EE}, and low-\(\ell\) \textit{TT}, \textit{EE}, and PR4 lensing data \cite{2020c, Efstathiou_2021, Carron_2022a, Carron_2022b, Rosenberg_2022}; BAO -- DESI data \cite{adame2024desi}; and SN -- Pantheon+ data \cite{Scolnic_2022}. The data combination of CMB, BAO, and SN in this paper is also abbreviated as CBS.

	The equation of state (EoS) of dark energy is given by \( w = p/\rho \), where \( p \) is the pressure and \( \rho \) is the energy density. We employ different models to describe its dynamics. (i) $\Lambda$CDM: \( w = -1 \). (ii) \( w \)CDM: \( w(z) = w \), where \( w \) is a constant EoS for dark energy. (iii) \( w_0w_a \)CDM: \( w \) varies with redshift according to the parameterization \( w(z) = w_0 + w_az / (1 + z) \), where \( w_0 \) is the present-day value of the EoS parameter and \( w_a \) describes its evolutionary behavior.
        \begin{figure}[t]
       \centering
        \includegraphics[width=0.9\textwidth]{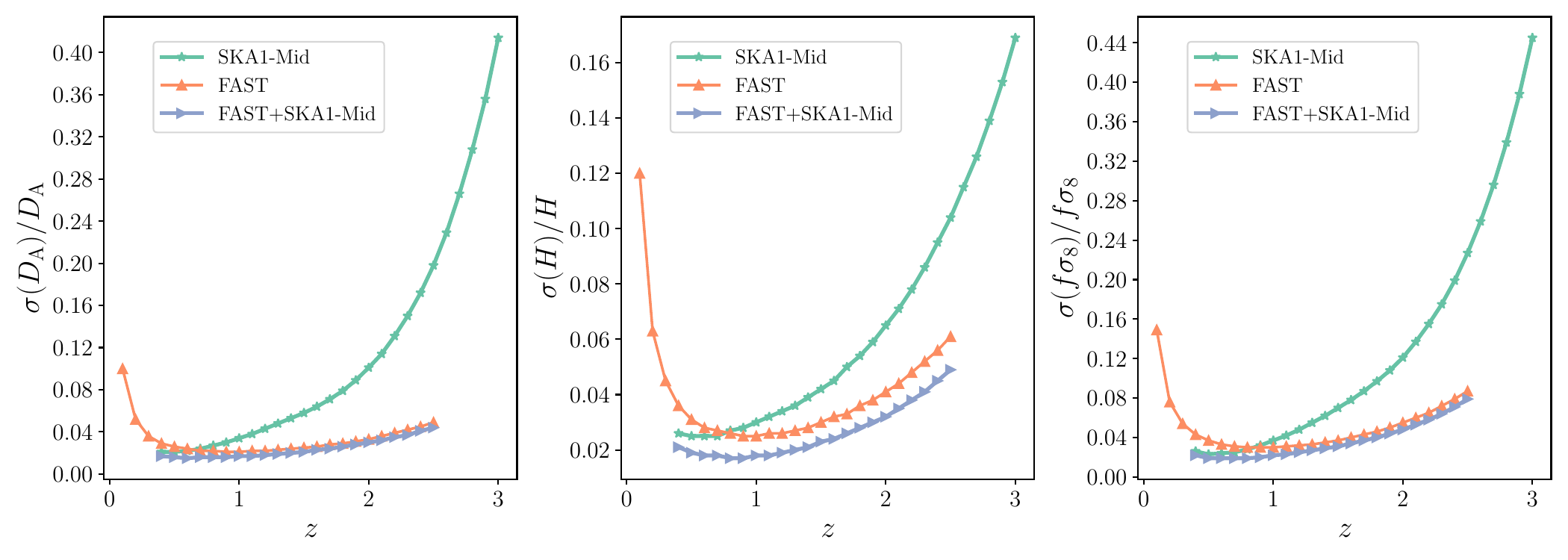}
        \caption{Relative errors of \(D_\mathrm{A}(z)\), \(H(z)\), and \(f\sigma_8(z)\) for 21 cm IM with FAST, SKA1-Mid, and FAST+SKA1-Mid.}
        \label{SKA1}
    \end{figure}
    
	\section{Results and discussion}\label{sec3}
	In this section, we  first discuss the capabilities of various instruments and their combinations in 21 cm IM using the relative errors of BAO parameters \(\{\varepsilon(D_\mathrm{A}(z)), \varepsilon(H(z)), \varepsilon(f\sigma_8(z))\}\), where \(\varepsilon(\xi) = \sigma(\xi)/\xi\). We then present the cosmological parameter constraints employing FAST, SKA1-Mid, FASTA, and CBS, followed by comparison and discussion of the results.

\subsection{Combination of FAST and SKA-Mid}\label{sec3.2}
In this subsection, we discuss the capabilities of SKA1-Mid and FAST with a wide-band receiver, as well as their combination in 21 cm IM. Due to geographical factors, the overlapping observable sky area for FAST and SKA1-Mid is very limited, which necessitates a strategy where each telescope observes different sky regions.

{It should be noted that for SKA1-Mid, we consider the single dish mode. This is because, in interferometer mode, the maximum observable scale is determined by the shortest baseline. Although SKA1-Mid has an advantage in angular resolution in interferometer mode, it lacks very short baselines, which limits its ability to observe large angular scales. Its sparse array design results in low baseline density and does not achieve good uv coverage, making it unsuitable for 21 cm IM. Improving its performance in interferometer mode would require a denser antenna array, like the CHIME design.
In single dish mode, the maximum observable scale is determined by the aperture of the telescope, allowing SKA1-Mid to observe larger scales. Single dish mode also offers other advantages: a wider field of view, more beams, enabling broader sky coverage within limited observation time. Therefore, single dish mode is an optimal choice for 21 cm IM on cosmological scales \cite{bull2015late, Zhang:2023gaz}.}

From Figure~\ref{SKA1}, it is evident that FAST and SKA1-Mid have distinct advantages across different redshift ranges. SKA1-Mid achieves tight constraints within the range of \(0.35 < z < 0.8\) due to its high survey efficiency. In this range, the relative errors in BAO parameters are typically below 3\% (see also Ref.~\cite{wu2022prospects}). On the other hand, FAST attains robust constraints within the range of \(0.8 < z < 2.5\) because of its high transverse resolution and sensitivity. In this range, the relative errors in BAO parameters are typically around or below 4\%.

Their combined observations enable precise measurement of BAO parameters across the redshift range of \(0 < z < 2.5\), achieving constraints on \(D_\mathrm{A}(z)\), \(H(z)\), and \(f\sigma_8(z)\) to below 2\% for most of this range, and to as low as 1\% at \(z \approx 1\).

Upgrading the existing receiver to detect 21 cm signals up to \(0 < z < 2.5\) will enable FAST to utilize its high-redshift advantage in 21 cm IM. Combined with SKA1-Mid, this complementary strategy will improve measurements of BAO parameters across a wide redshift range.
    \subsection{Combination of FAST and 40-m antennas}\label{sec3.1}
	In this subsection, we discuss the extent to which incorporating 40-m antennas with FAST enhances the constraints on BAO parameters in 21 cm IM. An array of 40-m antennas, located in close geographical proximity to FAST, enhances the survey efficiency by enabling observations of the same sky regions.

\begin{figure}[t]
        \centering
        \includegraphics[width=0.9\textwidth]{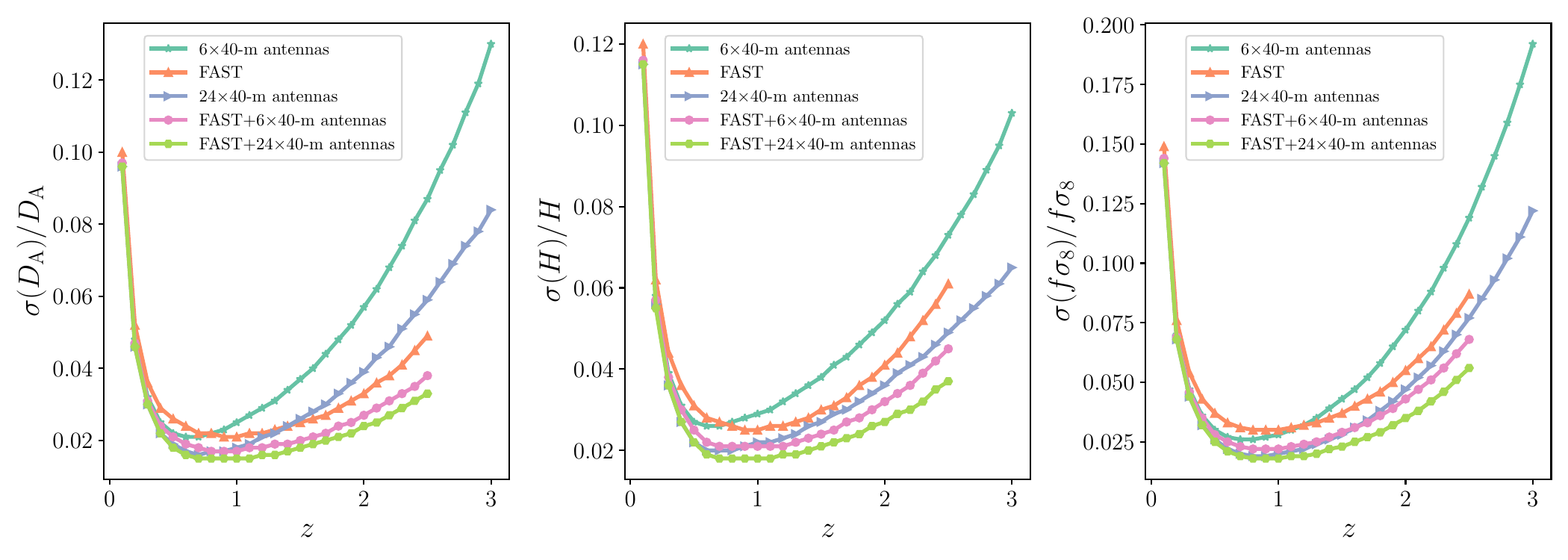} 
        \caption{Relative errors of $D_\mathrm{A}(z)$, $H(z)$, and $f\sigma_8(z)$ for 21 cm IM with FAST, 40-m antennas, FAST+$6\times 40$-m antennas, and FAST+$24\times 40$-m antennas.}
        \label{fig:6 Dishes} 
    \end{figure}

Figure~\ref{fig:6 Dishes} illustrates the fractional constraints achievable on \(D_\mathrm{A}(z)\), \(H(z)\), and \(f\sigma_8(z)\) using 21 cm IM with an array of 40-m antennas, FAST, and their combined configuration. The 40-m antennas perform better at low redshifts because they have higher survey efficiency. FAST is more effective at higher redshifts due to its greater sensitivity and resolution. The combination of FAST and 40-m antennas significantly reduces observational errors, particularly at high redshifts. The relative errors in BAO parameters are typically below 4\%, with errors reaching as low as 2\% at \(z \approx 1\).

For the configurations of $6\times 40$-m antennas and $24\times 40$-m antennas combined with FAST, Figure~\ref{fig:6 Dishes} shows that the larger array provides only marginal improvements. This result suggests that the combination of FAST with $6\times 40$-m antennas offers a more optimal configuration for the 21 cm IM survey.

	\begin{table*}[t]
        \caption{The errors for the cosmological parameters in the \(\Lambda\)CDM, $w$CDM, and $w_{0}w_{a}$CDM models using CBS and 21 cm IM. Note that here \(H_0\) is in units of km s\(^{-1}\) Mpc\(^{-1}\).
		}
    
            \renewcommand\arraystretch{1.4}
		\centering
		\resizebox{\textwidth}{!}{
			\begin{tabular}{lccccccccc}
				\hline\hline
				\multirow{2}{*}{Data} & \multicolumn{2}{c}{\underline{\hspace{1.3cm} {$\Lambda \mathrm{CDM_{\,}}$}\hspace{1cm}}} & \multicolumn{3}{c}{\underline{\hspace{2cm} {$w \mathrm{CDM_{\,}}$}\hspace{2cm}}} & \multicolumn{4}{c}{\underline{\hspace{3.1cm} {$w_{0}w_{a}$CDM}\hspace{3.1cm}}} \\
				& $ \sigma(\Omega_\mathrm{m})/10^{-3} $ & $ \sigma(H_0)/10^{-1} $ & $ \sigma(\Omega_\mathrm{m})/10^{-3} $ & $ \sigma(H_0)/10^{-1} $ & $ \sigma(w)/10^{-2} $ & $ \sigma(\Omega_\mathrm{m})/10^{-3} $ & $ \sigma(H_0)/10^{-1} $ & $ \sigma(w_0)/10^{-2} $ & $ \sigma(w_a)/10^{-1} $ \\ 
				\hline
				CBS    & 4.6 & 3.4 & 6.5 & 6.4 & 2.4 & 6.4 & 6.5 & 6.8 & 3.0 \\
				SKA1-Mid   & 8.7 & 7.0 & 9.0 & 9.1 & 4.5 & 26.0 & 15.0 & 16.0 & 7.3 \\
				FAST   & 7.6 & 5.1 & 7.8 & 7.2 & 3.6 & 21.0 & 14.0 & 14.0 & 5.7 \\
				FASTA & 3.6 & 2.6 & 3.7 & 3.9 & 2.0 & 12.0 & 8.9 & 9.0 & 3.3 \\
				\hline
			\end{tabular}
		}
		\label{sigma}
	\end{table*}
 
    \begin{figure}[t]
    \centering
    \resizebox{0.9\textwidth}{!}{
    \begin{subfigure}{0.25\textwidth}
        \centering
        \includegraphics[width=\textwidth]{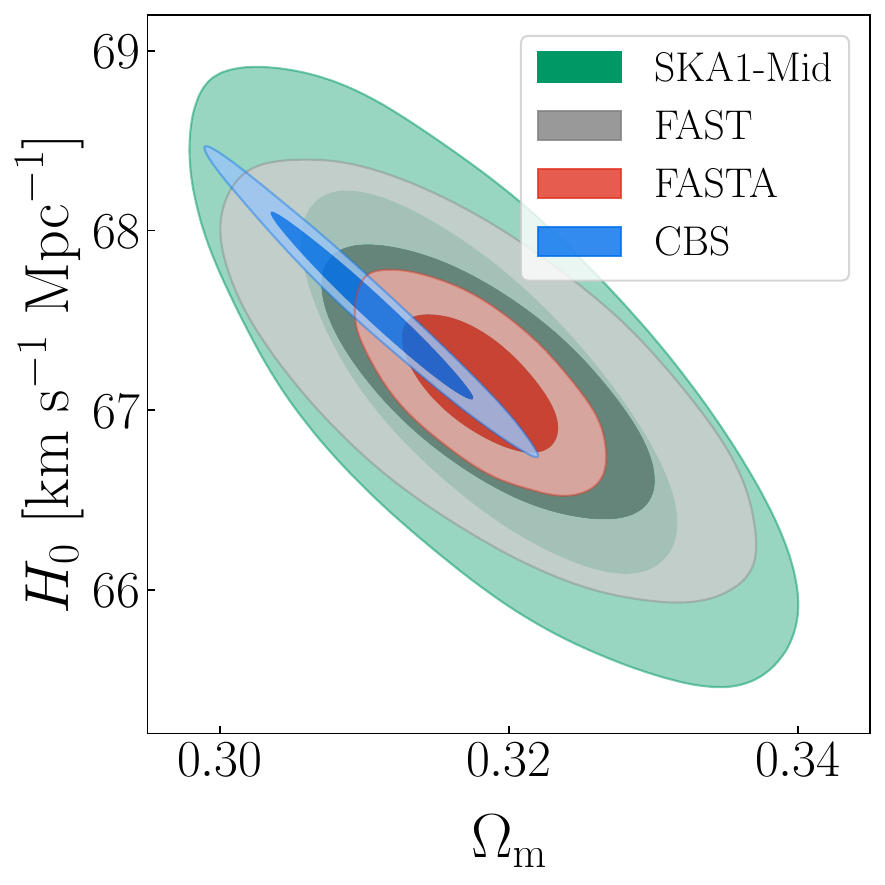}
    \end{subfigure}
    \begin{subfigure}{0.25\textwidth}
        \centering
        \includegraphics[width=\textwidth]{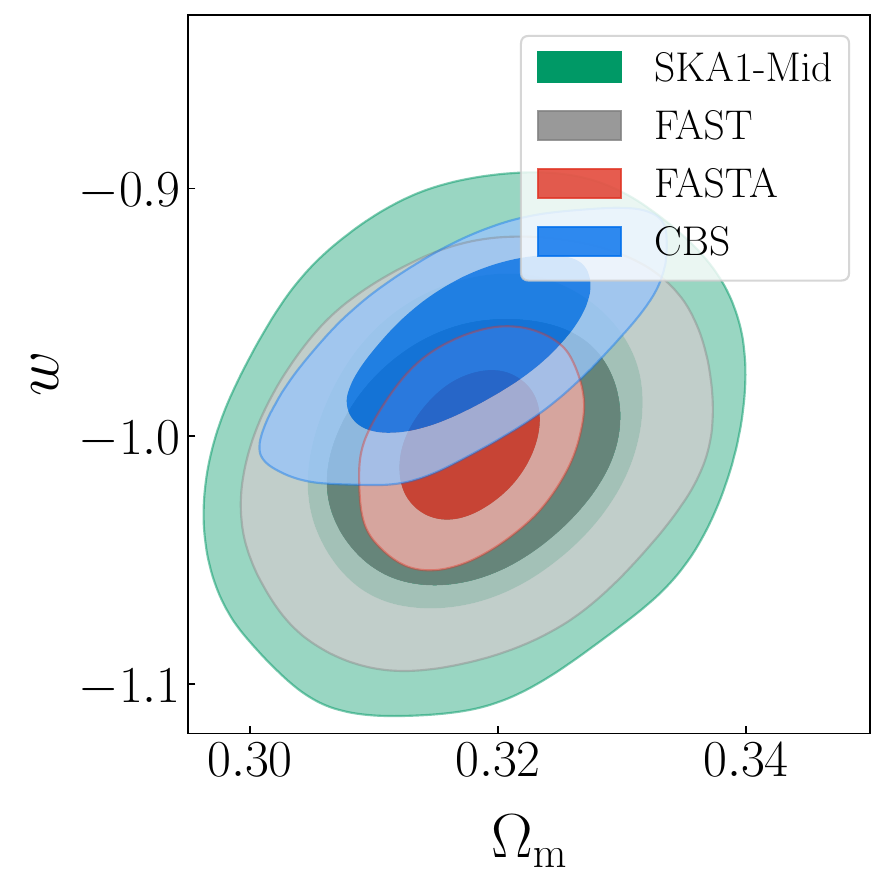}
    \end{subfigure}
    \begin{subfigure}{0.2625\textwidth}
        \centering
        \includegraphics[width=\textwidth]{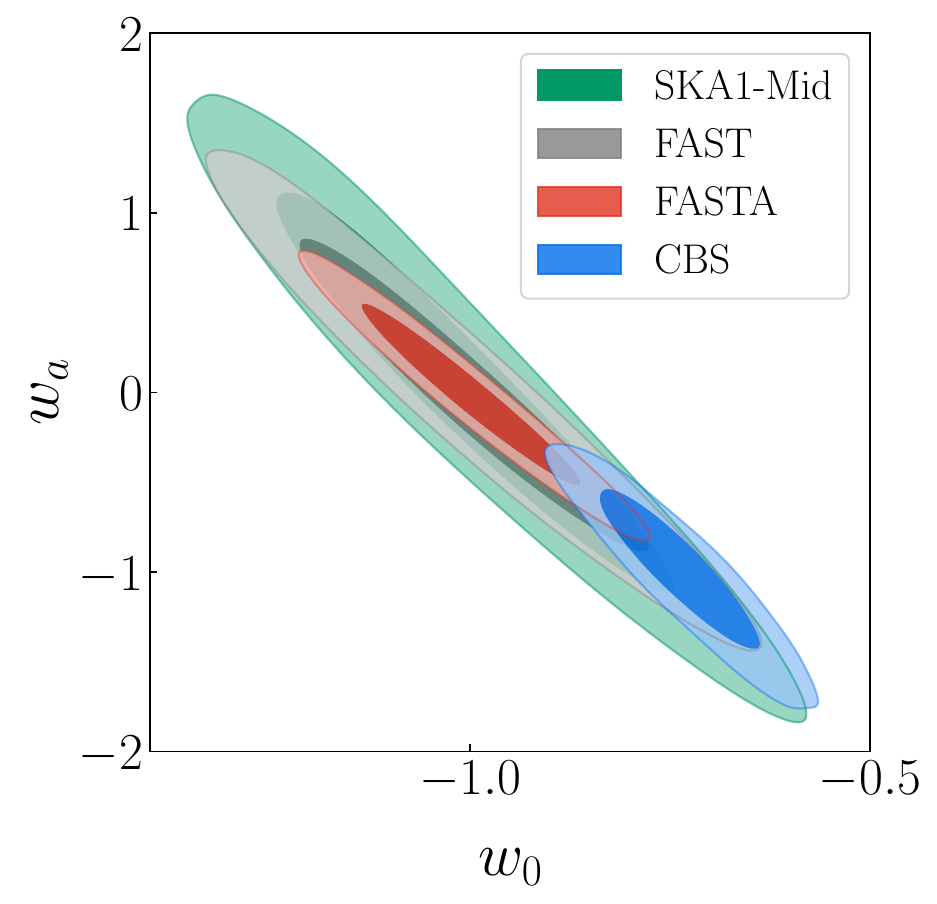}
    \end{subfigure}
    }
    \caption{Constraints ({68.3\% and 95.4\%} confidence level) on the dark energy models using simulated data of 21 cm IM with SKA1-Mid, FAST, FASTA, and observational data of CBS. Left, middle, and right panels show the constraints on the $\Lambda$CDM, $w$CDM, and $w_{0}w_{a}$CDM models, respectively.
}
    \label{con} 
\end{figure}    
    \subsection{FASTA}\label{sec3.3}
	In this subsection, we present the cosmological parameters constraints by using simulations of 21 cm IM data with different instruments and current observational data from CBS. Figure~\ref{con} shows the  parameter constraints within the {68.3\% and 95.4\%} confidence intervals for the $\Lambda$CDM, $w$CDM, and $w_0w_a$CDM models. The specific values are listed in Table~\ref{sigma}.

Compared to the results in Ref.~\cite{wu2022prospects} for FAST ($0 < z < 0.35$), a wide-band receiver proposed in this paper utilizes the resolution advantage of the FAST at high redshifts, significantly enhancing the constrains on cosmological parameters. The main specific parameter constraints are as follows: for the $\Lambda$CDM model, FAST achieves $\sigma(\Omega_\mathrm{m}) = 0.0076$ and $\sigma(H_0) = 0.51 \mathrm{~km} \mathrm{~s}^{-1} \mathrm{~Mpc}^{-1}$. For the $w$CDM model, the constraint is $\sigma(w) = 0.036$. For the $w_0w_a$CDM model, the constraints are $\sigma(w_0) = 0.14$, and $\sigma(w_a) = 0.57$. Compared to SKA1-Mid, FAST also has higher sensitivity and lower receiver noise, which enables it to achieve better parameter constraints. For the dark-energy EoS parameters, there is an improvement of approximately 20\% for $w$, 12.5\% for $w_0$, and 21.9\% for $w_a$.

\begin{figure*}[t]
        \centering
        \includegraphics[width=0.9\textwidth]{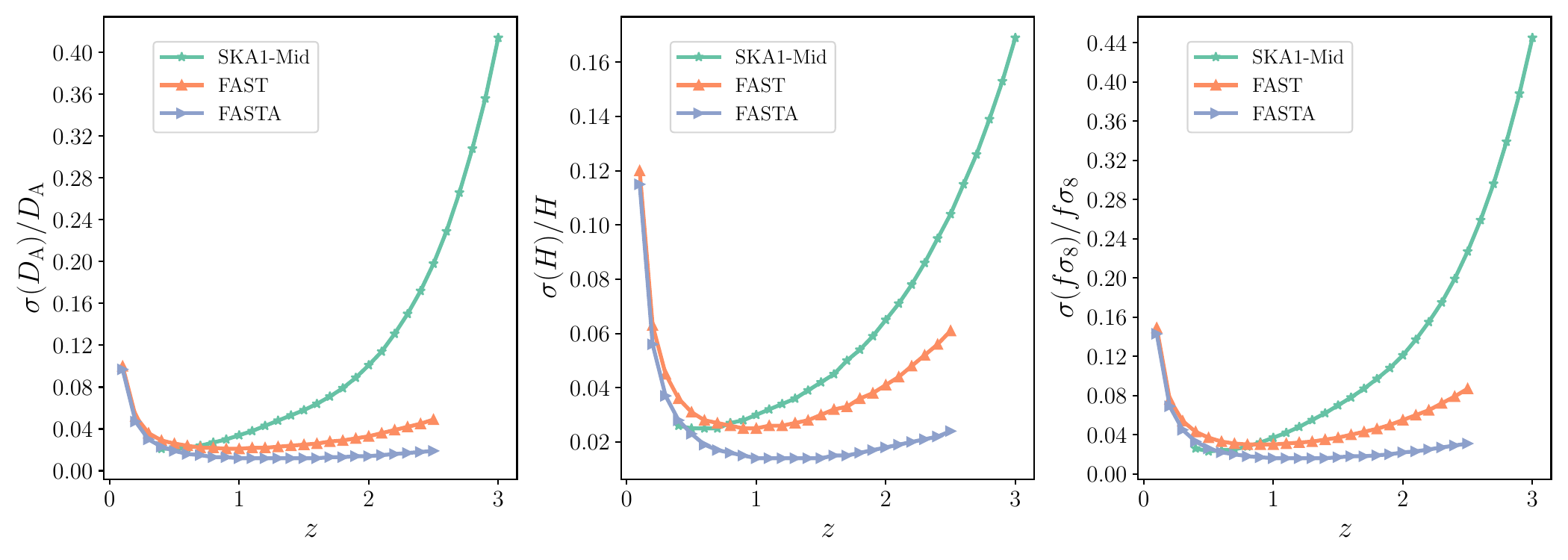} 
        \caption{Relative errors of $D_\mathrm{A}(z)$, $H(z)$, and $f\sigma_8(z)$ for 21 cm IM with SKA1-Mid, FAST, and FASTA.}
        \label{fig:FASTA} 
    \end{figure*}

	Further advances are observed with FASTA. From Figure~\ref{fig:FASTA}, FASTA performs well in observing BAO parameters across the entire band, achieving an observational error as low as 1\%. This precise observation of BAO parameters translates into tight constraints on cosmological parameters. 
 
In the $\Lambda$CDM model, FASTA provides best constraints, \(\sigma(\Omega_m) = 3.6 \times 10^{-3}\) and \(\sigma(H_0) = 0.26 \mathrm{~km} \mathrm{~s}^{-1} \mathrm{~Mpc}^{-1}\), which surpasses other instruments and CBS.
In the $w$CDM model, FASTA also provides best constraints, \(\sigma(\Omega_\mathrm{m}) = 0.0037\), \(\sigma(H_0) = 0.39 \mathrm{~km} \mathrm{~s}^{-1} \mathrm{~Mpc}^{-1}\), and \(\sigma(w) = 0.02\). For the parameter \(w\), FASTA offers an improvement of approximately 55.56\% compared to SKA1-Mid, 44.44\% compared to FAST, and 16.67\% compared to CBS.
In the $w_0w_a$CDM model, FASTA achieves highly precise parameter constraints, with \(\sigma(\Omega_\mathrm{m}) = 0.012\), \(\sigma(H_0) = 0.89 \mathrm{~km} \mathrm{~s}^{-1} \mathrm{~Mpc}^{-1}\), \(\sigma(w_0) = 0.09\), and \(\sigma(w_a) = 0.33\). For \(w_0\) and \(w_a\), FASTA provides an improvement of approximately 43.75\% and 54.79\% compared to SKA1-Mid, and 35.71\% and 42.11\% compared to FAST, while achieving constraints comparable to CBS using 21 cm IM  alone.

In summary, FASTA significantly improves cosmological parameter constraints across various models, surpassing other instruments and achieving results comparable to CBS using 21 cm IM  alone.
    \begin{figure*}[t]
        \centering
        \includegraphics[width=0.9\textwidth]{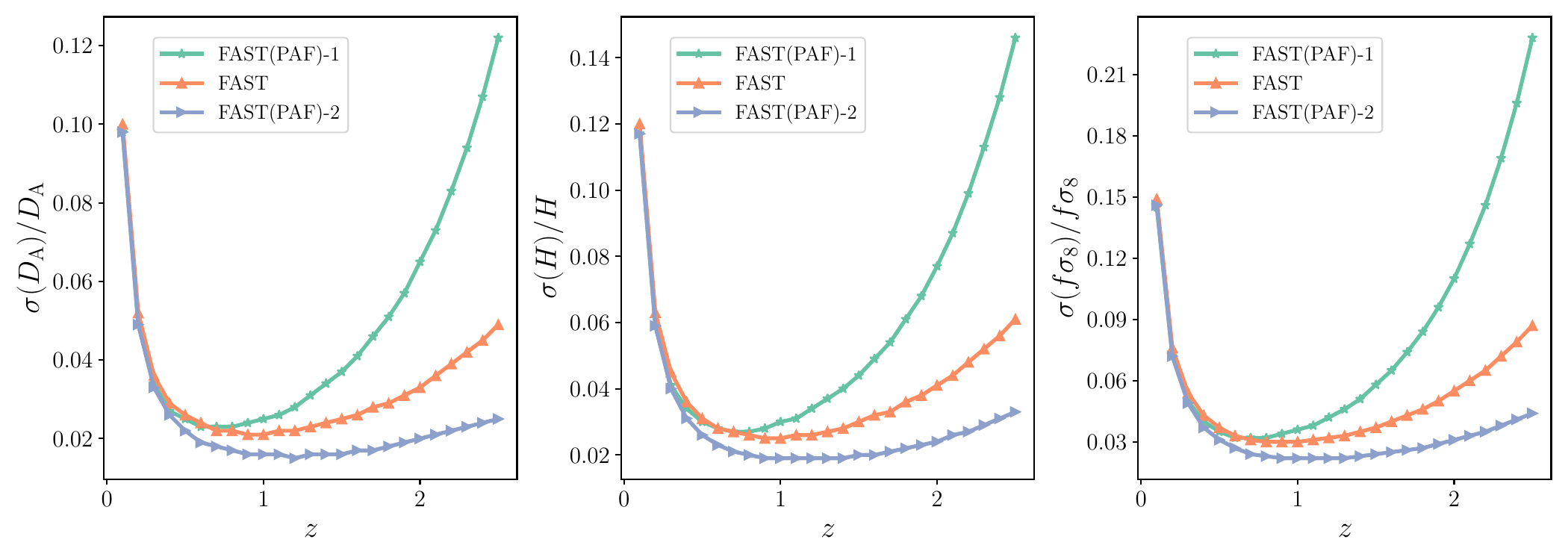}
        \caption{Relative errors of \(D_\mathrm{A}(z)\), \(H(z)\), and \(f\sigma_8(z)\) for 21 cm IM with FAST and FAST(PAF), where FAST(PAF)-1 represents the fixed FOV scenario and FAST(PAF)-2 represents the fixed $N_\mathrm{b}$ scenario.}
        \label{PAF} 
    \end{figure*}

	\subsection{Phased array feed}\label{sec3.4}
	In this subsection, we discuss the enhancements and challenges of upgrading the feed of FAST to PAF for 21 cm IM. PAF can achieve various beamforming modalities by adjusting the phase and amplitude of each element in the array, thereby generating multiple beams simultaneously. This capability covers a larger field of view ($\mathrm{FOV}$) and enables rapidly beam steering and scanning \cite{landon2010phased, hotan2021australian}.  At the same time, different beamforming strategies significantly impact noise characteristics, and data calibration and processing.  We considered two beamforming strategies in terms of survey efficiency.
 
    In the first scenario, we consider a fixed $\mathrm{FOV}$ and maximize the $N_\mathrm{b}$ within this field. The size of a beam is given by $\theta = {c}/(\nu D_\mathrm{dish}) $. If there is no overlap between the beams, the maximum $N_\mathrm{b}$ can be written as
    \begin{equation}
    N_\mathrm{b} = \mathrm{FOV} \cdot \left(\frac{\nu D}{c}\right)^2 .
    \end{equation}

    In this case, \( C^\mathrm{N} \) follows the same pattern as described in Ref.~\cite{bull2015late}
    \begin{equation}
		C^\mathrm{N}(\boldsymbol{q}, y) \rightarrow C^\mathrm{N}(\boldsymbol{q}, y) \times \begin{cases}1, & \nu>\nu_{\text {crit }} \\ \left(\nu_{\text {crit }} / \nu\right)^2, & \nu \leqslant \nu_{\text {crit }}\end{cases} .
    \end{equation}
	Here we have $\nu_\mathrm{crit}=1420\ \mathrm{MHz}$. As shown in Figure~\ref{PAF}, the errors increase significantly at low frequencies. This is because at low frequencies, the beam size increases, resulting in a decrease in \( N_\mathrm{b} \).
 
	Another scenario is a fixed $N_\mathrm{b}$. We assume the case of 100 beams. As shown in Figure~\ref{PAF}, the increase in $N_\mathrm{b}$ significantly enhances survey efficiency, achieving best constraint result of 2\% around \( z = 1 \). However, in practice, the $\mathrm{FOV}$ does not increase endlessly as the beam size increases at lower frequencies. As the $\mathrm{FOV}$ certainly deviates from the boresight, the survey efficiency decreases. This requires actual measurements to PAF, but here we make an ideal estimate. The real situation is expected to lie between the fixed $\mathrm{FOV}$ and the fixed $N_\mathrm{b}$ scenarios.

    \section{Conclusion}\label{sec4}
	In this paper, we explore several potential upgrades to FAST using 21 cm IM to enhance cosmological research. Those upgrades include FAST with a wide-band receiver, the combination of 40-m antennas and FAST, FASTA, and FAST with a PAF. We simulated the full 21 cm IM power spectrum to constrain the BAO and RSD parameters \(D_\mathrm{A}(z)\), \(H(z)\), and \(f\sigma_8(z)\), and used these results to evaluate the capabilities of these different configurations. Finally, we used the simulated BAO and RSD data to constrain the cosmological parameters of the \(\Lambda\)CDM, \(w\)CDM, and \(w_0w_a\)CDM models.

	Firstly, the wide-band receiver utilizes the capabilities of FAST in the redshift range \(0.8 < z < 2.5\), achieving better cosmological parameter constraints than SKA1-Mid. It also enables FAST to form a complementary observation with SKA1-Mid, which has strengths in the redshift range of \(0 < z < 0.8\), providing high-precision observations across a wider redshift range.
 
	Secondly, we discuss the combination of the 40-m antennas and FAST. Incorporating 40-m antennas with FAST significantly enhances survey efficiency, thereby reducing BAO parameter errors to typically below 2\%. For the combination with FAST, 6\(\times\)40-m antennas are a more balanced configuration compared to 24\(\times\)40-m antennas for 21 cm IM.
	
	Next, we discuss the cosmological parameter constraints for FASTA. Through BAO parameter measurements approaching 1\% accuracy, it shows excellent performance in constraining cosmological parameters. Specifically, in the $w_0w_a$CDM model, FASTA achieves highly precise parameter constraints, with improvements of 43.75\% for \(w_0\) and 54.79\% for \(w_a\) over SKA1-Mid, and 35.71\% for \(w_0\) and 42.11\% for \(w_a\) over FAST, achieving results comparable to CBS using 21 cm IM  alone.

	Lastly, we analyze FAST with a PAF. We consider two beamforming strategies, a fixed FOV and a fixed $N_\mathrm{b}$. The real situation is likely to fall between these two cases. This provides a basis for understanding the advantages and limitations of different beamforming strategies for 21 cm IM. A balanced beamforming strategy will be beneficial for effectively utilizing PAF.
	
	In summary, these upgrades to FAST have enormous potential to improve the precision of cosmological measurements. Despite facing some technical challenges, they can greatly enhance our understanding of the evolution of the universe and the nature of dark energy.

	\begin{acknowledgments}
	We thank Guang-Chen Sun, Ji-Guo Zhang,  Zi-Qiang Zhao, and Lu Feng for helpful discussions and suggestions. We are grateful for the support from the National SKA Program of China (Grants Nos. 2022SKA0110200 and 2022SKA0110203), the National Natural Science Foundation of China (Grants Nos. 12473001, 11975072, 11875102, 11835009, and 12305069), the National 111 Project (Grant No. B16009), and the Program of the Education Department of Liaoning Province (Grant No. JYTMS20231695).

	\end{acknowledgments}

\bibliographystyle{JHEP}
\bibliography{paper}
 
 \end{document}